\PassOptionsToPackage{unicode}{hyperref}
\PassOptionsToPackage{hyphens}{url}
\PassOptionsToPackage{dvipsnames,svgnames,x11names}{xcolor}
\documentclass[
]{article}
\usepackage{amsmath,amssymb}
\usepackage{lmodern}
\usepackage{iftex}
\ifPDFTeX
  \usepackage[T1]{fontenc}
  \usepackage[utf8]{inputenc}
  \usepackage{textcomp} 
\else 
  \usepackage{unicode-math}
  \defaultfontfeatures{Scale=MatchLowercase}
  \defaultfontfeatures[\rmfamily]{Ligatures=TeX,Scale=1}
\fi
\IfFileExists{upquote.sty}{\usepackage{upquote}}{}
\IfFileExists{microtype.sty}{
  \usepackage[]{microtype}
  \UseMicrotypeSet[protrusion]{basicmath} 
}{}
\makeatletter
\@ifundefined{KOMAClassName}{
  \IfFileExists{parskip.sty}{%
    \usepackage{parskip}
  }{
    \setlength{\parindent}{0pt}
    \setlength{\parskip}{6pt plus 2pt minus 1pt}}
}{
  \KOMAoptions{parskip=half}}
\makeatother
\usepackage{xcolor}
\usepackage{graphicx}
\makeatletter
\def\maxwidth{\ifdim\Gin@nat@width>\linewidth\linewidth\else\Gin@nat@width\fi}
\def\maxheight{\ifdim\Gin@nat@height>\textheight\textheight\else\Gin@nat@height\fi}
\makeatother
\setkeys{Gin}{width=\maxwidth,height=\maxheight,keepaspectratio}
\makeatletter
\def\fps@figure{htbp}
\makeatother
\newlength{\cslhangindent}
\newlength{\csllabelwidth}
\newlength{\cslentryspacingunit} 
\newenvironment{CSLReferences}[2] 
 {
  \setlength{\parindent}{0pt}
  \ifodd #1
  \let\oldpar\par
  \def\par{\hangindent=\cslhangindent\oldpar}
  \fi
  \setlength{\parskip}{#2\cslentryspacingunit}
 }%
 {}
\usepackage{calc}

\ifLuaTeX
\usepackage[bidi=basic]{babel}
\else
\usepackage[bidi=default]{babel}
\fi
\babelprovide[main,import]{american}

\def\languageshorthands#1{}
\ifLuaTeX
  \usepackage{selnolig}  
\fi
\IfFileExists{bookmark.sty}{\usepackage{bookmark}}{\usepackage{hyperref}}
\IfFileExists{xurl.sty}{\usepackage{xurl}}{} 
\hypersetup{
  pdftitle={`votess: A multi-target, GPU-capable, parallel Voronoi
tessellator'},
  pdfauthor={Samridh Dev Singh, Chris Byrohl, Dylan Nelson},
  pdflang={en-US},
  colorlinks=true,
  linkcolor={Maroon},
  filecolor={Maroon},
  citecolor={Blue},
  urlcolor={Blue},
  pdfcreator={LaTeX via pandoc}}

\title{`votess: A multi-target, GPU-capable, parallel Voronoi
tessellator'}

\usepackage[affil-it]{authblk}
\usepackage{orcidlink}
\author[1%
  ]{Samridh Dev Singh%
    \,\orcidlink{0009-0008-8620-3648}\,%
    }
\author[2%
  ]{Chris Byrohl%
    \,\orcidlink{0000-0002-0885-8090}\,%
    }
\author[2%
  ]{Dylan Nelson%
    \,\orcidlink{0000-0001-8421-5890}\,%
    }

\affil[1]{Grinnell College, 1115 8th Avenue, 50112 Grinnell, United
States of America}
\affil[2]{Heidelberg University, Institute for Theoretical Astronomy,
Albert-Ueberle-Str. 2, 69120 Heidelberg, Germany}
\date{15 September 2024}

\begin{document}
\maketitle

\hypertarget{statement-of-need}{%
\section{Statement of need}\label{statement-of-need}}

The Voronoi tessellation is a spatial decomposition that uniquely
partitions space into convex sub-regions based on proximity to a
discrete set of generating points. It is widely used across scientific
domains. In astrophysics, Voronoi meshes are used in observational data
analysis as well as numerical simulations of cosmic structure formation
(\protect\hyperlink{ref-Springel2010}{Springel, 2010}). The increasing
size of modern datasets has underscored the need for faster, more
efficient algorithms and numerical methods. The rise of powerful
many-core CPU and GPU architectures has greatly increased the available
computational power.

However, most existing implementations of Voronoi tessellations are
tailored to specific architectures, limiting their portability.
Additionally, classic sequential insertion algorithms have difficulty
using multi-core systems in a fully parallel manner. No general purpose,
publicly available code capable of GPU-accelerated parallel Voronoi mesh
generation in 3D space is available.

\hypertarget{summary}{%
\section{Summary}\label{summary}}

\texttt{votess} is a library for computing parallel 3D Voronoi
tessellations on heterogeneous platforms, from CPUs to GPUs to future
accelerator architectures. To do so, it uses the SYCL single-source
framework abstraction. \texttt{votess} was designed to be portable yet
performant, accessible to both developers and users with several
easy-to-use interfaces.

The underlying algorithm computes the Voronoi structure cell-by-cell
(\protect\hyperlink{ref-ray2018}{Ray et al., 2018}). It produces the
geometry of the Voronoi cells and their neighbor connectivity
information, rather than a full combinatorial mesh data structure. This
simplifies the method and makes it more ammenable to data parallel
architectures than alternatives such as sequential insertion or the
Bowyer-Watson algorithm.

The core method of \texttt{votess} consists of two main steps. First,
the input set of points is sorted into a grid, and a k-nearest neighbors
search is performed. Once the k nearest neighbors are identified for
each point, the Voronoi cell is computed by iteratively clipping a
bounding box using the perpendicular bisectors between the point and its
nearby neighbors. A ``security radius'' condition ensures that the
resulting Voronoi cell is valid. If a cell cannot be validated, an
automatic CPU fallback mechanism ensures robustness.

This efficient algorithm allows for independent thread execution, making
it highly suitable for multi-core CPUs as well as GPU parallelism. This
independence of cell computations achieves significant speedups in
parallel environments.

\hypertarget{performance}{%
\subsection{Performance}\label{performance}}

\includegraphics{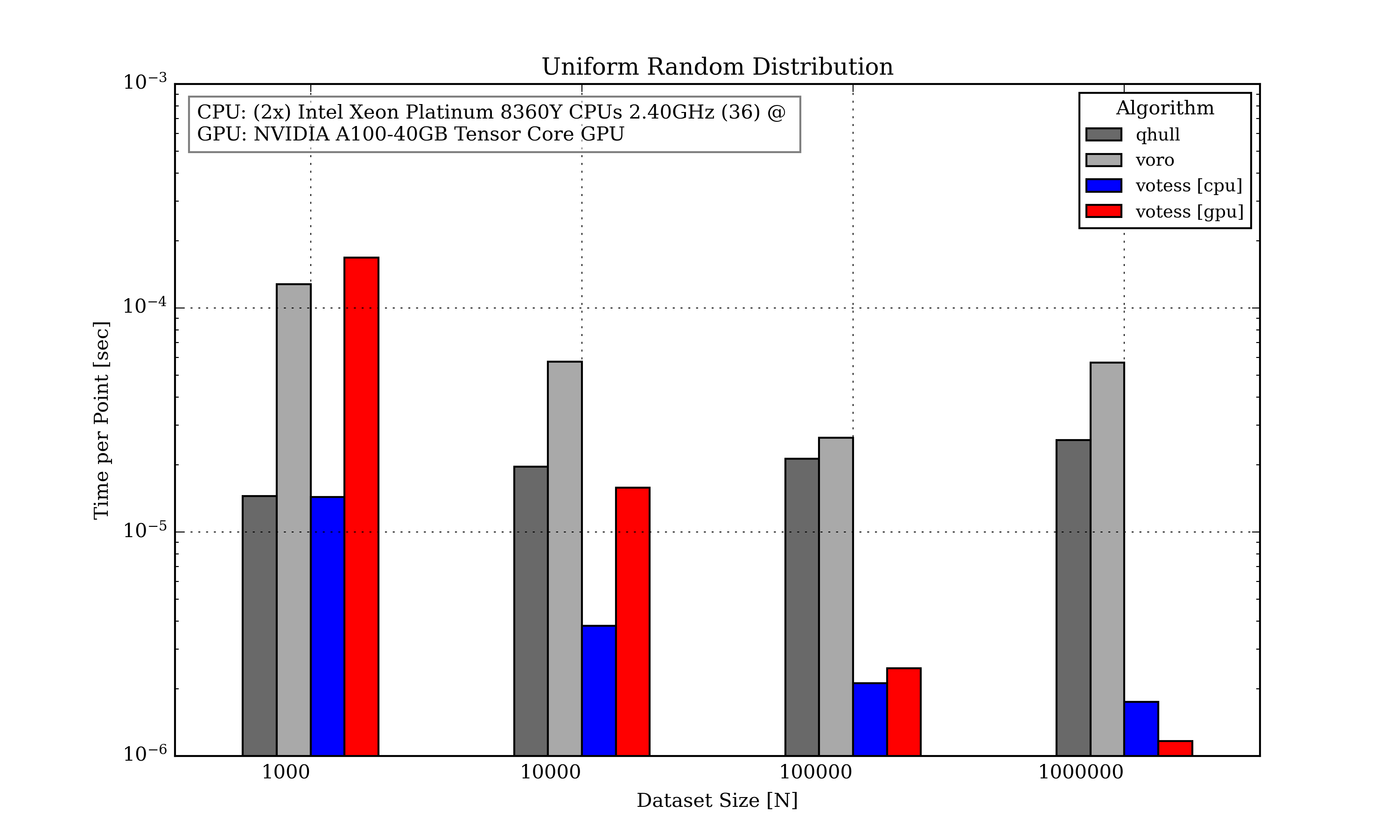}

As expected, \texttt{votess} significantly outperforms single-threaded
alternatives.

In Figure 1, we show its performance compared to two other
single-threaded Voronoi tessellation libraries: \texttt{Qhull} and
\texttt{Voro++}. Both are well-tested and widely used. \texttt{Qhull} is
a computational geometry library that constructs convex hulls and
Voronoi diagrams using an indirect projection method
(\protect\hyperlink{ref-10.1145ux2f235815.235821}{Barber et al., 1996}),
while \texttt{Voro++} is a C++ library specifically designed for
three-dimensional Voronoi tessellations, utilizing a cell-based
computation approach that is well-suited for physical applications
(\protect\hyperlink{ref-rycroft2009voro}{Rycroft, 2009}).

We find that \texttt{votess} performs best on GPUs with large datasets.
The CPU implementation can outperform other implementations by a factor
of 10 to 100.

Multithreaded Voronoi tesellelation codes do exist, and these include
\texttt{ParVoro++} (\protect\hyperlink{ref-WU2023102995}{Wu et al.,
2023}), \texttt{CGAL} (\protect\hyperlink{ref-cgal2018}{The CGAL
Project, 2018}), and \texttt{GEOGRAM}
(\protect\hyperlink{ref-geogram2018}{Inria, 2018}). However, they do not
natively support GPU architectures.

\hypertarget{features}{%
\section{Features}\label{features}}

\texttt{votess} is designed to be versatile. It supports various
outputs, including the natural neighbor information for each Voronoi
cell. This is a 2D jagged array of neighbor indices of the sorted input
dataset.

Users can invoke \texttt{votess} in three ways: through the C++ library,
a command-line interface \texttt{clvotess}, and a Python wrapper
interface \texttt{pyvotess}. The C++ library offers a simple interface
with a \texttt{tessellate} function that computes the mesh. The Python
wrapper, mirrors the functionality of the C++ version, with native numpy
array support, providing ease of use for Python-based workflows.

The behavior of \texttt{votess} can be fine-tuned with run time
parameters in order to (optionally) optimize runtime performance.

\hypertarget{acknowledgements}{%
\section{Acknowledgements}\label{acknowledgements}}

CB and DN acknowledge funding from the Deutsche Forschungsgemeinschaft
(DFG) through an Emmy Noether Research Group (grant number NE 2441/1-1).

\hypertarget{references}{%
\section*{References}\label{references}}
\addcontentsline{toc}{section}{References}

\hypertarget{refs}{}
\begin{CSLReferences}{1}{0}
\leavevmode\vadjust pre{\hypertarget{ref-10.1145ux2f235815.235821}{}}%
Barber, C. B., Dobkin, D. P., \& Huhdanpaa, H. (1996). The quickhull
algorithm for convex hulls. \emph{ACM Trans. Math. Softw.},
\emph{22}(4), 469--483. \url{https://doi.org/10.1145/235815.235821}

\leavevmode\vadjust pre{\hypertarget{ref-geogram2018}{}}%
Inria, P. A.-G. (2018). \emph{Geogram: A programming library of
geometric algorithms}.
\url{http://alice.loria.fr/software/geogram/doc/html/index.html}

\leavevmode\vadjust pre{\hypertarget{ref-ray2018}{}}%
Ray, N., Sokolov, D., Lefebvre, S., \& Lévy, B. (2018). Meshless voronoi
on the GPU. \emph{ACM Transactions on Graphics}, \emph{37}(6), 1--12.
\url{https://doi.org/10.1145/3272127.3275092}

\leavevmode\vadjust pre{\hypertarget{ref-rycroft2009voro}{}}%
Rycroft, C. (2009). \emph{VORO++: A three-dimensional voronoi cell
library in c++}.

\leavevmode\vadjust pre{\hypertarget{ref-Springel2010}{}}%
Springel, V. (2010). Moving-mesh hydrodynamics with the AREPO code.
\emph{Proceedings of the International Astronomical Union},
\emph{6}(S270), 203--206.
\url{https://doi.org/10.1017/S1743921311000378}

\leavevmode\vadjust pre{\hypertarget{ref-cgal2018}{}}%
The CGAL Project. (2018). \emph{{CGAL} user and reference manual}
(4.12.1 ed.). {CGAL Editorial Board}.
\url{https://doc.cgal.org/4.12.1/Manual/packages.html\#PkgSpatialSortingSummary}

\leavevmode\vadjust pre{\hypertarget{ref-WU2023102995}{}}%
Wu, G., Tian, H., Lu, G., \& Wang, W. (2023). ParVoro++: A scalable
parallel algorithm for constructing 3D voronoi tessellations based on
kd-tree decomposition. \emph{Parallel Computing}, \emph{115}, 102995.
https://doi.org/\url{https://doi.org/10.1016/j.parco.2023.102995}

\end{CSLReferences}

\end{document}